\documentclass[twocolumn,secnumarabic,amssymb, amsmath, nofootinbib,tightenlines,
nobibnotes, aps, prl,epsfig]{revtex4}
\usepackage{graphicx}
\usepackage{dcolumn}
\usepackage{bm}
\begin{document}
\preprint{APS/123-QED}
\title{The study of the gluon distribution function and reduced cross section behavior using the proton structure function }

\author{B.Rezaei }
\altaffiliation{brezaei@razi.ac.ir}

\author{G.R.Boroun}%
 \email{grboroun@gmail.com; boroun@razi.ac.ir }

\affiliation{ Physics Department, Razi University, Kermanshah
67149, Iran}
\date{\today}
\begin{abstract}
The behavior of the gluon distribution function and the reduced
cross section considered from the proton structure function and
its derivatives
 at low values of $x$. These behaviors studied
and compared with the experimental data. These results are
augmented by including an additional higher-twist term in the
description of the nonlinear correction. This additional term,
modified nonlinear correction, improves the description of the
reduced cross sections significantly at low values of $Q^{2}$. We
discuss, furthermore, how this behavior can be determine the
reduced cross section with respect to the proton parameterization
at high-y values. The resulting predictions for $\sigma_{r}$
suggest that further corrections are required for $Q^{2}$ less than about $3~\mathrm{GeV}^{2}$.\\
\end{abstract}
 \pacs{***}
\maketitle
\subsection{Introduction}
The gluon distribution  in hadrons at small Bjorken $x$ play a
vital role in our understanding of the deep-inelastic scattering
(DIS) and our predictions for new physics at the Large Hadron
Collider (LHC) in ultra-high energy processes. It is important to
know the gluon distribution inside a hadron at small $x$ because
gluons are expected to be dominant in this region. The gluons
couple only through the strong interaction. Consequently the
gluons cannot be measured directly in the DIS. Indeed, the gluon
density is much higher than of charged partons and the
photoabsorbtion will be dominated by the boson-gluon fusion (BGF)
$\gamma^{*}p{\rightarrow}q\overline{q}$ [1].\\
The high-luminosity LHC program would be uniquely complemented by
the proposed Large Hadron electron Collider (LHeC), where it is a
high-energy lepton-proton and lepton-nucleus collider based at
CERN [2]. The kinematic range in the ($x,Q^{2}$) plane of the LHeC
for electron and positron neutral-current (NC) for the high energy
$E_{p}=7~\mathrm{TeV}$  is $5{\times}10^{-6}\leq x \leq 0.8$ and
$5 \leq Q^{2} \leq 10^{6}~\mathrm{GeV}^{ 2}$ and $5 \leq Q^{2}
\leq 5 {\times}10^{5}~\mathrm{GeV}^{ 2}$ respectively. Today, an
integrated Future Circular Collider programme consisting of a
luminosity-frontier highest-energy lepton collider followed by an
energy-frontier hadron collider is called FCC [3]. In this
collider  the FCC-eh with $50~ \mathrm{TeV}$ proton beams
colliding with 60 GeV electrons from an energy-recovery linac
would generate $\sim 2 \mathrm{ab}^{-1}$ of $3.5~ \mathrm{TeV}~
\mathrm{ep}$ collisions. Deep inelastic scattering measurements at
FCC-eh will allow the determination of the gluon saturation
phenomena required to unitarise the high-energy cross sections.
The determination of the high gluon densities and non-linear
dynamics at very small $x$ is  relevant to the interactions of
cosmic ultra high energy neutrinos. For HERA data, the gluon
density is not determined at low $x$. But, with the FCC-eh, a
precision of a few percent at small $x$ becomes possible down to
nearly $x{\simeq}10^{-6}$ [3].\\
The gluon and quark distributions are determined mainly by the
proton structure function $F_{2}^{\gamma p}(x,Q^{2})$ measured in
ep DIS. The longitudinal structure function and the slope of the
transverse structure function become direct probes of the gluon
and sea quark distributions, over a large domain of values of $x$
and $Q^{2}$ (hereafter $Q^{2}$ is the virtuality of the photon).
Due to its origin, $F_{L}$ is directly sensitive to the gluon
distribution in the proton and consequently it is an important
quantity [4]. In perturbative quantum chromodynamics (pQCD) the
heavy quark production at HERA proceeds is dominantly via the
direct BGF. Therefore, once the distribution of the gluon inside
the proton is known, the heavy quark distribution can be easy
calculated from it. This process can be created when the squared
invariant mass of the hadronic final state has the condition the
runs as follows $W^{2}\geq 4m^{2}_{\mathrm{Heavy-quark}}$. At very
low $x$, the parton density cannot grow forever because hadronic
cross-sections comply with the unitary bound known as Froissart
Bound [5]. For this region the gluon recombination is known to
provide the mechanism responsible for the unitarization of the
cross section at high energies. In other words, the multiple
gluon-gluon interactions provide nonlinear corrections in the
DGLAP evolution equations [6]. While the precise measurement of
$F_{L}$ at the FCC-eh and  $F_{2}$ can discover whether gluon
saturates.\\
We know that the DGLAP evolution equations are the basic
perturbative tools to study the $Q^{2}$ evolution of parton
distribution functions (PDFs) in the deep inelastic regime. The
solutions of the coupled DGLAP evolution equations will provide
the gluon and singlet quark distributions inside the nucleon at
small $x$. These distribution functions have been determined
simultaneously by starting with a virtuality $Q_{0}^{2}\leq
(m_{c}^{2}{\approx}2~ \mathrm{GeV^{2}})$ [7]. The initial
distributions for the parton distributions usually are  determined
in a global QCD analysis including a wide range of DIS data from
collider  until fixed
 experiments. In Ref.[8], a new form of the DIS structure function (SF) $F_{2}(x,Q^{2})$
 in the domain $10^{-3} \leq x \leq 0.09$ and $0.11~ \mathrm{GeV^{2}}\leq Q^{2} \leq 1200~ \mathrm{GeV^{2}}$ of
  HERA data was proposed. This new form of the proton structure function leads to the small $x$ asymptotics of the reduced
cross-sections $\sim \ln^{2}1/x$, which is in turn in an agreement
with the Froissard predictions [5]. The modified form of the
proton structure function with a new parametrization which
describes fairly well the available experimental data,  at
asymptotically small $x$, have been presented in Refs.[9-10]. This
parametrization suggested is  pertinent in investigations of
lepton-hadron processes at ultra-high energies, which obtained
from a combined fit of the H1 and ZEUS collaborations data [11] in
a range of the kinematical variables $x$ and $Q^{2}$, $x \leq 0.1$
and $0.15~ \mathrm{GeV^{2}} \leq Q^{2} \leq 3000~ \mathrm{GeV^{2}}$.\\
In Ref.[12], an analytical relation has been derived for
calculating the gluon distribution function within the
Laplace-transform method at leading order. The gluon distribution
is determined by the parametrization of $F_{2}$ (Ref.[8]) as a
simultaneous function of $x$ and $Q^{2}$ in a domain
$x_{min}{\leq}x{\leq}x_{max}$ and
$Q^{2}_{min}{\leq}Q^{2}{\leq}Q^{2}_{max}$. Then, in Ref.[13],
authors extended those previous derivation of the gluon
distribution at leading order to include the effects of
heavy-quark mass. These analytical solutions of the gluon behavior
using a differntial-equation method [12] and a Laplace transform
technique [13] have been reported with considerable
phenomenological success at leading order (LO) approximation
respectively. Also the solutions of the unpolarized DGLAP equation
for the QCD evolution of gluon distribution have been discussed
considerably over the past years [14,15].  In two recent papers
[16-17] the behavior of the gluon distribution function
$G(x,Q^{2})$ studied at small values of $x$ by using a global
parameterization of the data on $F_{2}(x,Q^{2})$ at LO analysis.\\
In the present paper we study the behavior of the gluon
distribution function using the proton structure function and its
derivatives with respect to ${\ln}Q^{2}$ in high-order corrections
based on the hard-pomeron exchange [4,18]. Then we estimate the
nonlinear corrections to the gluon distribution function behavior
at small values of $x$. Therefore, to obtain a precise evidence of
the nonlinear corrections in the HERA kinematic region, we
consider the longitudinal structure function that directly depends
on the behavior of the gluon distribution in which the active
quarks are treated as massless. Then we generalize to the case of
reduced cross section using the same method. We use the modified
nonlinear corrections and higher twist predictions in obtaining
our analytical solutions at low $Q^{2}$ and low $x$. A completed
comparison between the obtained results at NNLO approximation and
available DIS data are presented.\\

\subsection{Analytical treatment of the gluon distribution function}

The DGLAP evolution equation for the singlet quark structure
function is given by
\begin{eqnarray}
\frac{{\partial}F_{2}^{s}(x,Q^{2})}{{\partial}{\ln}Q^{2}}&=&\frac{\alpha_{s}}{2\pi}{\int_{x}^{1}}dz[
P_{qq}(z,\alpha_{s}(Q^{2}))
F_{2}^{s}(\frac{x}{z},Q^{2})\nonumber\\
&&+2N_{f}P_{qg}(z,\alpha_{s}(Q^{2})) G(\frac{x}{z},Q^{2})]
\end{eqnarray}
where  $F_{2}^{s}(x,Q^{2})$ and $G(x,Q^{2})$ are singlet and gluon
distribution functions. Here the representation for the gluon
distribution $G(x,Q^{2})=xg(x,Q^{2})$ is used where $g(x,Q^{2})$
is the gluon density . The splitting functions $P_{ij}^{,}s$ are
the LO, NLO and NNLO Altarelli- Parisi splitting kernels as
\begin{eqnarray}
P_{ij}(x,\alpha_{s}(Q^{2}))&=&P_{ij}^{\rm
LO}(x)+\frac{\alpha_{s}(Q^{2})}{2\pi}P_{ij}^{\rm
NLO}(x)\nonumber\\
&&+(\frac{\alpha_{s}(Q^{2})}{2\pi})^{2} P_{ij}^{\rm NNLO}(x),
\end{eqnarray}
which the explicit forms of the splitting functions at LO upto
NNLO are given in Refs.[19-20].\\
The running coupling constant $\alpha_{s}$ has the following forms
in LO, NLO and NNLO approximation respectively [21]
\begin{equation}
\alpha_{s}^{\rm LO}=\frac{4\pi}{\beta_{0}t},
\end{equation}
\begin{equation}
\alpha_{s}^{\rm
NLO}=\frac{4\pi}{\beta_{0}t}[1-\frac{\beta_{1}{\ln}t}{\beta_{0}^{2}t}],
\end{equation}
and
\begin{eqnarray}
\alpha_{s}^{\rm
NNLO}&=&\frac{4\pi}{\beta_{0}t}[1-\frac{\beta_{1}{\ln}t}{\beta_{0}^{2}t}+\frac{1}{(\beta_{0}t)^{2}}
[(\frac{\beta_{1}}{\beta_{0}})^{2}\nonumber\\
&&(\ln^{2}t-{\ln}t+1)+\frac{\beta_{2}}{\beta_{0}}]].
\end{eqnarray}
where $\beta_{0}=\frac{1}{3}(33-2n_{f})$,
$\beta_{1}=102-\frac{38}{3}n_{f}$ and
$\beta_{2}=\frac{2857}{6}-\frac{6673}{18}n_{f}+\frac{325}{54}n_{f}^{2}$.
The variable $t$ is defined as
$t={\ln}(\frac{Q^{2}}{\Lambda^{2}})$ and $\Lambda$ is the QCD
cut- off parameter at each heavy quark mass threshold as we take the $n_{f}=4$ for $m_{c}^{2}<\mu^{2}<m^{2}_{b}$.\\
The power law behavior of the singlet and gluon distribution
functions introduced as $F_{2}^{s}{\sim}x^{-\lambda_{s}}$ and
$G{\sim}x^{-\lambda_{g}}$ [4,18] where  exponents $\lambda_{s}$
and $\lambda_{g}$ are given as the derivatives: $
\lambda_{s}=\frac{\partial \ln F_{2}^{s}(x,Q^{2})}{\partial
\ln(1/x)}$ and $ \lambda_{g}=\frac{\partial \ln
G(x,Q^{2})}{\partial \ln(1/x)}$. By considering the hard-power
behavior of the singlet and gluon distribution functions, one can
rewrite the evolution equation in terms the convolution integrals,
using the convolution symbol $\otimes$, as
\begin{eqnarray}
\frac{{\partial}F_{2}(x,Q^{2})}{{\partial}{\ln}Q^{2}}&=&F_{2}(x,Q^{2})\Phi_{qq}(x,Q^{2})\nonumber\\
&&+G(x,Q^{2})\Theta_{qg}(x,Q^{2}),
\end{eqnarray}
where $F_{2}=\frac{5}{18}F_{2}^{s}$ and the nonsinglet
contribution is negligible at small $x$ and can be ignored. The
kernels for the quark and gluon sectors, denoted by $\Phi$ and
$\Theta$, presented at LO up to NNLO respectively as
\begin{eqnarray}
\Theta_{qg}(x,Q^{2})&=&P_{qg}(x,\alpha_{s}){\odot}
x^{\lambda_{g}},\nonumber\\
\Phi_{qq}(x,Q^{2})&=&P_{qq}(x,\alpha_{s}){\odot}
x^{\lambda_{s}},\nonumber\\
\end{eqnarray}
where  $ f(x){\odot}g(x){\equiv}\int_{x}^{1}(dy/y)f(y)g(y)$. The
singlet exponent $\lambda_{s}$ is found to be $\simeq 0.33$ in
Refs.[22-23] and  the gluon exponent $\lambda_{g}$ is found to be
$\simeq 0.43$ [4,18,23]. Therefore the gluon distribution function
is defined by the following form
\begin{eqnarray}
G(x,Q^{2})=
\frac{1}{\Theta_{qg}(x,Q^{2})}[\frac{{\partial}F_{2}(x,Q^{2})}{{\partial}{\ln}Q^{2}}
-\Phi_{qq}(x,Q^{2})F_{2}(x,Q^{2})],\nonumber\\
\end{eqnarray}
where the parameterization $F_{2}(x,Q^{2})$ and its derivatives
suggested in Refs.[8-10]. This result is  general and simple and
gives the leading order upto high-order expressions for the gluon
distribution once the parameterization $F_{2}$ is known. This
relation primarily comes from the extension of range and precision
in the measurements of $F_{2}(x,Q^{2})$ and
$\frac{{\partial}F_{2}(x,Q^{2})}{{\partial}{\ln}Q^{2}}$, which at
small $x$ are measures of the gluon density.\\
We write our final analytical answer by the following form as
\begin{widetext}
\begin{eqnarray}
G(x,Q^{2})&=&\frac{1}{\{P_{qg}(x,\alpha_{s}){\odot}
x^{\lambda_{g}}\}}[\frac{{\partial}F_{2}(x,Q^{2})}{{\partial}{\ln}Q^{2}}
-\{P_{qq}(x,\alpha_{s}){\odot} x^{\lambda_{s}}\}F_{2}(x,Q^{2})].
\end{eqnarray}
\end{widetext}
The gluon distribution function at leading-order analysis is given
by
\begin{widetext}
\begin{eqnarray}
G^{\mathrm{LO}}(x,Q^{2})&=&\frac{\frac{9\pi}{5\alpha_{s}}}{\int_{x}^{1}{(z^2+(1-z)^2)z^{\lambda_{g}}dz}}[\frac{{\partial}F_{2}(x,Q^{2})}{{\partial}{\ln}Q^{2}}
-(\frac{\alpha_{s}}{4\pi}\{4+\frac{16}{3}\ln(\frac{1-x}{x})+\frac{16}{3}\int_{x}^{1}{\frac{z^{\lambda_{s}}-{z}^{-1}}{1-z}dz}\nonumber\\
&&-\frac{8}{3}\int_{x}^{1}{(1+z)z^{\lambda_{s}}dz\}})F_{2}(x,Q^{2})].
\end{eqnarray}
\end{widetext}
 The relations in Eqs.(8) until (10) give the leading
order and high-order expression for gluon distribution function in
terms of a function determined by the proton structure function.\\

{\bf Further consideration}\\

We now discuss how the presented results give the independent
evolution equation for the singlet structure function at small
$x$. The DGLAP evolution equation for the gluon distribution is
given by
\begin{eqnarray}
\frac{{\partial}G(x,Q^{2})}{{\partial}{\ln}Q^{2}}&=&G(x,Q^{2})\Phi_{gg}(x,Q^{2})\nonumber\\
&&+F_{2}(x,Q^{2})\Theta_{gq}(x,Q^{2}),
\end{eqnarray}
where kernels are
\begin{eqnarray}
\Theta_{gq}(x,Q^{2})&=&P_{gq}(x,\alpha_{s}){\odot}
x^{\lambda_{s}},\nonumber\\
\Phi_{gg}(x,Q^{2})&=&P_{gg}(x,\alpha_{s}){\odot}
x^{\lambda_{g}}.\nonumber\\
\end{eqnarray}
Let us consider the evolution of the gluon distribution
$G(x,Q^{2})$(Eq.8) with respect to ${\ln}Q^{2}$:\\

{\bf For fixed coupling case}\\

\begin{eqnarray}
\frac{{\partial}G(x,Q^{2})}{{\partial}{\ln}Q^{2}}&=&
\frac{1}{\Theta_{qg}(x,Q^{2})}[\frac{{\partial}^{2}F_{2}(x,Q^{2})}{{\partial}{\ln}^{2}Q^{2}}\nonumber\\
&&-\Phi_{qq}(x,Q^{2})\frac{{\partial}F_{2}(x,Q^{2})}{{\partial}{\ln}Q^{2}}].
\end{eqnarray}
After some rearranging a homogeneous second-order differential
equation is found. It is determined the proton structure function
$F_{2}(x,Q^{2})$ without having knowledge in terms of the gluon
distribution function, as
\begin{widetext}
\begin{eqnarray}
\frac{{\partial}^{2}F_{2}(x,Q^{2})}{{\partial}{\ln}^{2}Q^{2}}-[\Phi_{qq}(x,Q^{2})+\Phi_{gg}(x,Q^{2})]\frac{{\partial}F_{2}(x,Q^{2})}{{\partial}{\ln}Q^{2}}
+[\Phi_{qq}(x,Q^{2})\Phi_{gg}(x,Q^{2})-\Theta_{qg}(x,Q^{2})\Theta_{gq}(x,Q^{2})]F_{2}(x,Q^{2})=0\nonumber\\
\end{eqnarray}
\end{widetext}

{\bf For running coupling case}\\

The evolution of the gluon distribution $G(x,Q^{2})$(Eq.8) with
respect to ${\ln}Q^{2}$ is given by
\begin{widetext}
\begin{eqnarray}
\frac{{\partial}G(x,Q^{2})}{{\partial}{\ln}Q^{2}}&=&
\frac{{\partial}}{{\partial}{\ln}Q^{2}}(\frac{1}{\Theta_{qg}(x,Q^{2})})[\frac{{\partial}F_{2}(x,Q^{2})}{{\partial}{\ln}Q^{2}}
-\Phi_{qq}(x,Q^{2})F_{2}(x,Q^{2})]+
\frac{1}{\Theta_{qg}(x,Q^{2})}[\frac{{\partial}^{2}F_{2}(x,Q^{2})}{{\partial}{\ln}^{2}Q^{2}}\nonumber\\
&&-\frac{{\partial}\Phi_{qq}(x,Q^{2})}{{\partial}{\ln}Q^{2}}F_{2}(x,Q^{2})-\Phi_{qq}(x,Q^{2})\frac{{\partial}F_{2}(x,Q^{2})}{{\partial}{\ln}Q^{2}}].
\end{eqnarray}
\end{widetext}
In this case the evolution equation for the singlet structure
function decoupled takes the form
\begin{widetext}
\begin{eqnarray}
&&\frac{{\partial}^{2}F_{2}(x,Q^{2})}{{\partial}{\ln}^{2}Q^{2}}-[\frac{{\partial}{\ln}\Theta_{qg}(x,Q^{2})}{{\partial}{\ln}Q^{2}}+
\Phi_{qq}(x,Q^{2})+\Phi_{gg}(x,Q^{2})]\frac{{\partial}F_{2}(x,Q^{2})}{{\partial}{\ln}Q^{2}}\nonumber\\
&&+[\Phi_{qq}(x,Q^{2})\frac{{\partial}{\ln}\Theta_{qg}(x,Q^{2})}{{\partial}{\ln}Q^{2}}-\frac{{\partial}\Phi_{qq}(x,Q^{2})}{{\partial}{\ln}Q^{2}}
+\Phi_{qq}(x,Q^{2})\Phi_{gg}(x,Q^{2})-\Theta_{qg}(x,Q^{2})\Theta_{gq}(x,Q^{2})]F_{2}(x,Q^{2})=0
\end{eqnarray}
\end{widetext}

\subsection{Solution for ${\sigma_{r}}(x,Q^{2})$ using a Froissart bounded structure function $F_{2}(x,Q^{2})$}

The reduced neutral current (NC) deep inelastic $e^{±}p$
scattering cross sections are given by a linear combination of
generalised structure functions. The proton structure functions
$F_{2}$ and $F_{L}$ are related to the $\gamma^{*}p$ cross
sections of longitudinally and transversely polarized photons
$\sigma_{L}$ and $\sigma_{T}$ as $F_{L}\propto \sigma_{L}$ and
$F_{T}\propto \sigma_{L}+\sigma_{T}$. In one photon exchange
approximation, at low and moderate $Q^{2}$ (i.e.,$
Q^{2}{\leq}M^{2}_{z}{\approx}800~\mathrm{GeV^{2}}$), the reduced
cross section  is defined into the transverse and longitudinal
structure functions, $F_{2}(x,Q^{2})$ and $F_{L}(x,Q^{2})$, by the
following form
\begin{eqnarray}
{\sigma_{r}}(x,Q^{2})=F_{2}(x,Q^{2})-\frac{y^{2}}{Y_{+}}F_{L}(x,Q^{2}).
\end{eqnarray}
Here $Y_{+}=1+(1-y)^2$, $y={Q^{2}}/{xs}$ is the inelasticity and
$s$ is the center-of-mass  squared energy of  incoming electrons
and protons. The contribution of the term containing the
longitudinal structure function $F_{L}$ is only significant for
values of $y$ larger than approximately 0.5.  Indeed, the
longitudinal structure function is predominant in the cross
section in case of scattering of cosmic neutrinos from hadrons
[24,14]. Thus, the longitudinal structure function at low $x$ will
be checked in high energy process such as the Large Hadron
electron Collider (LHeC) project which runs to beyond a TeV in
center-of-mass energy [1].\\
In perturbative QCD, the longitudinal structure function in terms
of the coefficient functions is given by [25]
\begin{eqnarray}
x^{-1}F_{L}=C_{L,ns}{\otimes}q_{ns}+<e^{2}>(C_{L,q}{\otimes}q_{s}+C_{L,g}{\otimes}g),
\end{eqnarray}
where the non-singlet quark distribution, $xq_{ns}$, become
negligibly small in comparison with the singlet and gluon
distribution functions, $xq_{n}$ and $xg$, at low values of $x$
and can be ignored. $<e^{k}>$ is the average of the charge $e^{k}$
for the active quark flavors,
$<e^{k}>=n_{f}^{-1}\sum_{i=1}^{n_{f}}e_{i}^{k}$. The average
squared charge for even $n_{f}$ is equal to $5/18$. The symbol
${\otimes}$ denotes a convolution according to the usual
prescription,
$f(x){\otimes}g(x)=\int_{x}^{1}\frac{dy}{y}f(y)g(\frac{x}{y})$.
The perturbative expansion of the coefficient functions can be
written as [26]
\begin{eqnarray}
C_{L,a}(\alpha_{s},x)=\sum_{n=1}a(t)^{n}c_{L,a}^{n}(x),
\end{eqnarray}
where $n$ is the order in the running coupling constant. The
running coupling constant in the high-loop corrections of the
above  equation is expressed entirely  thorough the variable
$a(t)$, as $a(t)=\frac{\alpha_{s}}{4\pi}$. The explicit expression
for the coefficient functions in LO up to NNLO are relegated in
Ref.[26].\\
Exploiting the small- $x$ behavior of the  distribution functions
according to the hard (Lipatov) Pomeron, then Eq.(18) can be
written as
\begin{eqnarray}
F_{L}(x,Q^{2})&=&F_{2}(x,Q^{2})I_{L,s}(x,Q^{2})+G(x,Q^{2})I_{L,g}(x,Q^{2}),\nonumber\\
\end{eqnarray}
where
$I_{L,s}(x,Q^{2})=C_{L,q}(\alpha_{s},x){\odot}x^{\lambda_{s}}~$
and
$I_{L,g}(x,Q^{2})=<e^{2}>C_{L,g}(\alpha_{s},x){\odot}x^{\lambda_{g}}$.
On this basis the reduced cross section  obtain  by employing the
parametrization of $F_{2}(x,Q^{2})$ as
 \begin{widetext}
\begin{eqnarray}
{\sigma_{r}}(x,Q^{2})&=&F_{2}(x,Q^{2})\{1-\frac{y^{2}}{Y_{+}}I_{L,s}(x,Q^{2})\}-\frac{y^{2}}{Y_{+}}I_{L,g}(x,Q^{2})G(x,Q^{2})[{\equiv}Eq.8]\nonumber\\
&&=F_{2}(x,Q^{2})\{1-\frac{y^{2}}{Y_{+}}(I_{L,s}(x,Q^{2})-\frac{\Phi_{qq}(x,Q^{2})}{\Theta_{qg}(x,Q^{2})}I_{L,g}(x,Q^{2}))\}
-\frac{{\partial}F_{2}(x,Q^{2})}{{\partial}{\ln}Q^{2}}\{\frac{y^{2}}{Y_{+}}\frac{I_{L,g}(x,Q^{2})}{\Theta_{qg}(x,Q^{2})}\}.
\end{eqnarray}
\end{widetext}
Therefore  an explicit equation for reduced cross section in terms
of the proton structure function parametrization and its
derivatives is obtained.\\

\subsection{Nonlinear and Higher Twist corrections}

{\bf Nonlinear corrections}:\\

The screening effects are provided by a multiple gluon interaction
which leads to the nonlinear terms in the derivation of the linear
DGLAP evolution equations. Therefore the standard linear DGLAP
evolution equations will have to be  modified in order to take the
nonlinear corrections into account. Indeed the origin of the
shadowing correction , in pQCD interactions, is primarily
considered as the gluon recombination ($g+g \rightarrow g$) which
is simply the inverse process of gluon
splitting ($g \rightarrow g+g$).\\
Gribov, Levin, Ryskin, Mueller and Qiu (GLR-MQ) [27] performed a
detailed study of these recombination processes. This  widely
known as the GLR-MQ equation and involves the two-gluon
distribution per unit area of the hadron. This equation predicts a
saturation behavior of the gluon distribution at very small $x$
[28]. A closer examination of the small $x$ scattering is
resummation powers of $\alpha_{s}\ln(1/x)$ where leads to the
$k_{T}$-factorization form [29]. In the $k_{T}$-factorization
approach the large logarithms $\ln(1/x)$ are relevant for the
unintegrated gluon density in a nonlinear equation.  Solution of
this equation develops a saturation scale where tame the gluon
density
behavior at low values of $x$ and this is an intrinsic characteristic of a dense gluon system.\\
Therefore one should consider the low- $x$ behavior of the singlet
distribution using the nonlinear GLR-MQ evolution equation. The
shadowing correction to the evolution of the singlet quark
distribution can be written as [30]
\begin{eqnarray}
\frac{{\partial}xq(x,Q^{2})}{{\partial}{\ln}Q^{2}}&=&\frac{{\partial}xq(x,Q^{2})}{{\partial}{\ln}Q^{2}}|_{DGLAP}\nonumber\\
&&-\frac{27\alpha_{s}^{2}}{160R^{2}Q^{2}} [xg(x,Q^{2})]^{2}.
\end{eqnarray}
Eq. (22) can be rewrite in a convenient form as
\begin{eqnarray}
\frac{{\partial}F_{2}(x,Q^{2})}{{\partial}{\ln}Q^{2}}=\frac{{\partial}F_{2}(x,Q^{2})}{{\partial}lnQ^{2}}|_{DGLAP}-
\frac{5}{18}\frac{27\alpha_{s}^{2}}{160R^{2}Q^{2}}\nonumber\\
{\times}[xg(x,Q^{2})]^{2}.
\end{eqnarray}
The first term is the standard DGLAP evolution equation and the
value of $R$ is the correlation radius between two interacting
gluons. It will be  of the order of the proton radius
$(R\simeq5\hspace{0.1cm} GeV^{-1})$, if the gluons are distributed
through the whole of proton, or much smaller
$(R\simeq2\hspace{0.1cm} GeV^{-1})$ if gluons are concentrated in
hot- spot within the proton.\\
Also there is another mechanism to prevent generation of the high
density gluon states, as this is well known the vacuum color
screening [31]. There is a transition between the nonperturbative
and perturbative domains. In the QCD vacuum, the non-perturbative
fields form structures with sizes $\sim R_{c} $ which it is
smaller than $\Lambda_{QCD}$. The short propagation length for
perturbative gluons is $R_{c}\sim 0.2-0.3~fm$. The gluon fusion
effect in non-linear regime controlled by the new dimensionless
parameter $\sim \frac{R_{c}^{2}}{8B}$ where $B$ is the
characteristic size of the interaction region as this parameter
can be defined by $\ln(x_{0}/x)$ and $r$ where $r^{2} \sim
Q^{-2}$. Authors in this reference (i.e., Ref.[31]) show that the
nonlinear effects leads to the logarithmically ratio as the
nonlinear/linear effects are proportional to
$R_{c}^{2}/8B(\ln(x_{0}/x),r^{2})\ln(Q^{2}R^{2}_{c})$.\\
 Combining Eqs. (8) and (23), one could consider the
nonlinear correction to the gluon distribution function as
\begin{eqnarray}
G(x,Q^{2})&=&
\frac{1}{\Theta_{qg}(x,Q^{2})}[\frac{{\partial}F_{2}(x,Q^{2})}{{\partial}{\ln}Q^{2}}
-\frac{5}{18}\frac{27\alpha_{s}^{2}}{160R^{2}Q^{2}}{\times}\nonumber\\
&&G^{2}(x,Q^{2})-\Phi_{qq}(x,Q^{2})F_{2}(x,Q^{2})],\nonumber\\
\end{eqnarray}
where
\begin{eqnarray}
G(x,Q^{2})+\frac{1}{\Theta_{qg}(x,Q^{2})}\frac{5}{18}\frac{27\alpha_{s}^{2}}{160R^{2}Q^{2}}G^{2}(x,Q^{2})=\nonumber\\
\frac{1}{\Theta_{qg}(x,Q^{2})}[\frac{{\partial}F_{2}(x,Q^{2})}{{\partial}{\ln}Q^{2}}-\Phi_{qq}(x,Q^{2})F_{2}(x,Q^{2})].
\end{eqnarray}
Eq.(25) is a second-order equation which can be solved as
\begin{eqnarray}
G(x,Q^{2})&=&\mathcal{F}(x,Q^{2})[1-\frac{\mathcal{A}(x,Q^{2})}{\Theta_{qg}(x,Q^{2})}\mathcal{F}(x,Q^{2})\nonumber\\
&&+2(\frac{\mathcal{A}(x,Q^{2})}{\Theta_{qg}(x,Q^{2})}\mathcal{F}(x,Q^{2}))^{2}\nonumber\\
&&-5(\frac{\mathcal{A}(x,Q^{2})}{\Theta_{qg}(x,Q^{2})}\mathcal{F}(x,Q^{2}))^{3}+....]\nonumber\\
&&=\mathcal{F}(x,Q^{2})[1-\mathcal{N}+2\mathcal{N}^{2}-5\mathcal{N}^{3}+....]\nonumber\\
&&=\mathcal{F}(x,Q^{2})[\mathcal{NLC}],
\end{eqnarray}
where
$\mathcal{NLC}=\frac{\mathcal{A}(x,Q^{2})}{\Theta_{qg}(x,Q^{2})}\mathcal{F}(x,Q^{2})$,
$\mathcal{A}(x,Q^{2})=\frac{5}{18}\frac{27\alpha_{s}^{2}}{160R^{2}Q^{2}}$
and $\mathcal{F}(x,Q^{2})=
\frac{1}{\Theta_{qg}(x,Q^{2})}[\frac{{\partial}F_{2}(x,Q^{2})}{{\partial}{\ln}Q^{2}}-\Phi_{qq}(x,Q^{2})F_{2}(x,Q^{2})].$
 Therefore the nonlinear corrections ($\mathcal{NLC}s$) to the reduced cross
section  are obtained by the following forms
\begin{widetext}
\begin{eqnarray}
{\sigma_{r}}(x,Q^{2})|_{NLC}&=&F_{2}(x,Q^{2})\{1-\frac{y^{2}}{Y_{+}}(I_{L,s}(x,Q^{2})
-\frac{\Phi_{qq}(x,Q^{2})}{\Theta_{qg}(x,Q^{2})}I_{L,g}(x,Q^{2})[\mathcal{NLC}])\}
-\frac{{\partial}F_{2}(x,Q^{2})}{{\partial}{\ln}Q^{2}}\{\frac{y^{2}}{Y_{+}}\frac{I_{L,g}(x,Q^{2})}{\Theta_{qg}(x,Q^{2})}\}[\mathcal{NLC}].\nonumber\\
\end{eqnarray}
\end{widetext}

{\bf Higher Twist corrections}:\\

Introduction of higher-twist terms is one possible way to extend
the DGLAP framework to low $Q^{2}$ values. Such terms have been
introduced at low-$x$ values since, for the kinematics of HERA,
low $Q^{2}$ is only accessed at low $x$. To better illustrate our
calculations at low $Q^{2}$, we added a higher twist term in the
description of the structure functions for
 HERA data on deep inelastic scattering at low $x$
and low $Q^{2}$ values. It can be clearly seen that our
predictions with respect to the higher twist (HT) analyses are
comparable with data at this region. The leading twist
perturbative QCD predictions of the structure function $F_{2}$
augment by a simple higher twist term such that
\begin{eqnarray}
F_{2}^{HT}&=&F_{2}^{DGLAP}(1+\frac{A_{2}}{Q^{2}}),\nonumber\\
\frac{{\partial}F_{2}^{HT}(x,Q^{2})}{{\partial}{\ln}Q^{2}}&=&\frac{{\partial}F_{2}^{DGLAP}(x,Q^{2})}{{\partial}{\ln}Q^{2}}(1+\frac{A_{2}}{Q^{2}})\nonumber\\
&&-F_{2}^{DGLAP}\frac{A_{2}}{Q^{2}}.
\end{eqnarray}
where $A^{HT}_{2}=0.12~\pm~0.07~GeV^{2}$ is a free parameter at
NNLO [32-33]. Using the HT terms in Eq.(28), we can evaluate the
HT corrections to the reduced cross section as
\begin{widetext}
\begin{eqnarray}
{\sigma_{r}}(x,Q^{2})|_{_{HT}}&=&F_{2}(x,Q^{2})(1+\frac{A_{2}}{Q^{2}})\{1-\frac{y^{2}}{Y_{+}}(I_{L,s}(x,Q^{2})-\frac{\Phi_{qq}(x,Q^{2})}{\Theta_{qg}(x,Q^{2})}I_{L,g}(x,Q^{2}))\}\nonumber\\
&&-[\frac{{\partial}F_{2}(x,Q^{2})}{{\partial}{\ln}Q^{2}}(1+\frac{A_{2}}{Q^{2}})-F_{2}(x,Q^{2})
\frac{A_{2}}{Q^{2}}]\{\frac{y^{2}}{Y_{+}}\frac{I_{L,g}(x,Q^{2})}{\Theta_{qg}(x,Q^{2})}\}.\nonumber\\
\end{eqnarray}
\end{widetext}
{\bf Modified Nonlinear (MNL) and Higher Twist (HT) corrections}:\\

To proceed further, we use  the higher twist corrections to the
nonlinear behavior and put them in nonlinear corrections as a
modified nonlinear correction ($\mathcal{MNLC}$) propose to
replace the $\mathcal{NLC}$ as $\mathcal{MNLC}\equiv
\mathcal{NLC}+HTC$. The modified NLC reads
\begin{widetext}
\begin{eqnarray}
\mathcal{MNLC}=\frac{\mathcal{A}(x,Q^{2})}{\Theta_{qg}^{2}(x,Q^{2})}
[\{\frac{{\partial}F_{2}(x,Q^{2})}{{\partial}{\ln}Q^{2}}(1+\frac{A_{2}}{Q^{2}})-F_{2}(x,Q^{2})\frac{A_{2}}{Q^{2}}\}-\Phi_{qq}(x,Q^{2})
\{F_{2}(x,Q^{2})(1+\frac{A_{2}}{Q^{2}}) \}].\nonumber
 \end{eqnarray}
 \end{widetext}
Following the same procedure, the modified nonlinear and higher
twist corrections to the reduced cross section can be estimated as
\begin{widetext}
\begin{eqnarray}
{\sigma_{r}}(x,Q^{2})|_{MNLC}&=&F_{2}(x,Q^{2})(1+\frac{A_{2}}{Q^{2}})\{1-\frac{y^{2}}{Y_{+}}(I_{L,s}(x,Q^{2})
-\frac{\Phi_{qq}(x,Q^{2})}{\Theta_{qg}(x,Q^{2})}I_{L,g}(x,Q^{2})\mathcal{MNLC})\}\nonumber\\
&&-[\frac{{\partial}F_{2}(x,Q^{2})}{{\partial}{\ln}Q^{2}}(1+\frac{A_{2}}{Q^{2}})-F_{2}(x,Q^{2})\frac{A_{2}}{Q^{2}}]\{\frac{y^{2}}{Y_{+}}\frac{I_{L,g}(x,Q^{2})}{\Theta_{qg}(x,Q^{2})}\}\mathcal{MNLC}.\nonumber\\
\end{eqnarray}
\end{widetext}
Therefore we observe that our analysis is based on the gluon and
reduced cross section where the structure function and its
derivatives are supposed to be known and determined from the
existing experimental data.\\

\subsection{Results and Discussion}

In this section, we shall present our results that have been
obtained for the gluon distribution  function $G(x,Q^{2})$ and
reduced cross section ${\sigma_{r}}(x,Q^{2})$ using the
hard-pomeron behavior of the parton distributions. To investigate
this, a general relation (i.e., Eq.(8)) between the gluon density
and proton structure function and also the logarithmic slops
$F_{2}$ obtained. Several methods of relating the $F_{2}$ scaling
violation to the gluon density at low $x$ have been suggested
previously [34-36]. Recently a similar connection between the
gluon density and $\partial F_{2}/\partial \ln Q^{2}$ in FCC-eh
[3] from the extension of range and precision in the measurement
is suggested. Eq.(8) shows that the gluon density directly is
related to $F_{2}$ and $\partial F_{2}/\partial \ln Q^{2}$. Figure
1 shows the coefficient $\frac{\Phi_{qq}}{\Theta_{qg}}$  as a
function of $Q^{2}$ for the two $x$ values $1 \times 10^{-6}$ and
$1 \times 10^{-3}$. The coefficient of $F_{2}$ is zero at NNLO
only at $Q^{2}\simeq 200~ GeV^{2}$ and $Q^{2}\simeq 30~ GeV^{2}$
at low ($1 \times 10^{-6}$) and moderate ($1 \times 10^{-3}$) $x$
values respectively. In what follows it is convenient to directly
use the gluon distribution behavior with respect to $F_{2}$ and
$\partial F_{2}/\partial \ln Q^{2}$. One can see that proton
structure function and its derivative are supposed to be known
with respect to the paremetrizations represented in Refs.[8-10].
These parameterization obtained from a combined fit of the H1 and
ZEUS collaborations data [11] in a range of the kinematical
variables $x<0.01$ and
$0.15~\mathrm{GeV^{2}}<Q^{2}<3000~\mathrm{GeV^{2}}$.\\
In Fig.2 the determined  gluon distribution function is shown in a
wide range of $Q^{2}$ values  for $x=1 \times 10^{-3}$ at LO upto
NNLO approximation. The leading order gluon distribution is
compared with the most  parametrizations suggested at LO in
Refs.[12-13]. The explicit forms of the gluon distributions are
given in Appendix A. These results are accompanied with errors due
to fit parametrizations of $F_{2}$. In Ref.[23], the effective
exponents for singlet and gluon densities have been derived which
they are closer to those defined by the color dipole model and
hard-pomeron exponents [37]. The predictions for  the
 gluon distribution function in the kinematic range where it has been determined by
M.M.Block collaboration [8-10, 12-13] computed and compared at low values of $x$.\\
In Ref. [10] a new parametrization of the SF $F_{2}(x,Q^{2})$
which describes fairly well the available experimental data on the
reduced cross sections have been suggested. With respect to this
new parameterization, we consider the gluon behavior  in a wide
range of $Q^{2}$ values at low $x$ in Fig.3. In this figure we
have compared our results at LO upto NNLO approximation with those
obtained by authors in Refs.[12-13] at LO approximation.\\
A detailed comparison has also been shown with the NNLO results
 from Block 2014 parameterization [10] and
depicted in figure 4. In this figure our result based on the hard
pomeron behavior have been presented for gluon distribution at low
and moderate $Q^{2}$ values ($Q^{2}=3.5, ~5,~8.5, ~12 ~
\mathrm{and}~20~\mathrm{GeV}^{2}$). These results have been
compared with the NNLO analysis of JR09 model [38] and clearly
show significant agreement over a wide range of $x$ and $Q^{2}$
values.\\
The nonlinear corrections (NLCs) to the gluon density are
considered in a wide range of $Q^{2}$ values for different values
of $x$, viz. $1{\times}10^{-3},~1{\times}10^{-5}$ and
$1{\times}10^{-7}$ respectively. In Fig.5, the effect of
nonlinearity in our results is investigated for
$R=2~\mathrm{GeV}^{-1}$ at NNLO approximation. One can see that
obtained nonlinear corrections  for gluon distribution function
are larger for very low $x$ values ($x<10^{-3}$) at low $Q^{2}$
values ($1<Q^{2}<10~\mathrm{GeV}^{2}$) then another domains. Fig.6
represent linear and nonlinear results of of gluon distribution
function for $R=2~\mathrm{GeV}^{-1}$ at NNLO that have been
computed from Eqs.(8) and (26) for $Q^{2}=3.5,~8.5$ and
$20~\mathrm{GeV}^{2}$ respectively. It would appear that the
effect of nonlinearity at low $x$ should observe for low and
moderate $Q^{2}$. The nonlinear corrections can be neglected at
large values of $Q^{2}$, so we expect that our results are
consistent in the kinematic region $x\leq 0.01$ and moderate
$Q^{2}$ with other results. From these figurs, it is observed that
the NNLO nonlinear corrections (NNLO+NLCs) show tamed behavior
compared with only NNLO
approximation.\\
In Fig.7 we display our NNLO results for $\sigma_{r}$ with respect
to the parameterization of $F_{2}$ [10] at different proton beam
energies $E_{p}$ (=920, 820 and 575 $\mathrm{GeV}$), relevant for
most H1 measurements [39-40], where the turnover at small $x$
becomes more pronounced at smaller energies because of the larger
values of $y$. The data are taken with positrons of energy
$E_{e}=27.6~\mathrm{GeV}$ corresponding to the center of mass
energies $\sqrt{s}=2\sqrt{E_{e}E_{p}}[\mathrm{GeV}]$. In Fig.8 the
NNLO predictions at $\sqrt{s}=300.9~\mathrm{GeV}$ are compared
with H1 data [39] as accompanied with total errors. These
small-$x$ predictions are fully compatible with the H1 data
presented in [39]. The modified nonlinear  and higher twist
corrections are depicted and compared with the linear dependence
in these figures. We have also performed an analysis to check the
sensitivity of the modified nonlinear and higher twist corrections
in reduced cross
section results.\\
 We have calculated the $Q^{2}$-dependence, at low $x$, of the
reduced cross section $\sigma_{r}(x,Q^{2})$ (i.e., Eq.(30)) in the
NNLO approximation. Results of calculations and comparison with
data of the H1-Collaboration [39] are presented in Fig.10, where
the
 solid line correspond to the extracted $\sigma_{r}$ in the NNLO
approximations. Calculations have been performed at fixed value of
the inelasticity  $y$, $y=0.675$, allowing the Bjorken variable
$x$ to vary in the interval ($3\times 10^{-5}<x<7\times 10^{-4}$)
when $Q^{2}$ varies in the interval
($1.5<Q^{2}<40~\mathrm{GeV}^{2}$). To illustrate better the
modified nonlinear and higher twist corrections at low $Q^{2}$
values, we have plotted MNL+HT predictions to the NNLO results in
Fig.10. It may lead to the better determinations of the reduced
cross section specially at
low $Q^{2}$.\\

\subsection{Summary and Conclusion}

We presented Eqs.(8-10) for the extraction of the gluon
distribution function $G(x,Q^{2})$  at low $x$ in the
leading-order upto next-to-next-to-leading order from the
$F_{2}(x,Q^{2})$ and its ${\ln}Q^{2}$ derivative. Then rendered
the results for the reduced cross section $\sigma_{r}(x,Q^{2})$ in
a wide range of $Q^{2}$ values. The reduced cross section behavior
is considered at low $x$ with respect to the $F_{2}$
parameterization. These results based on an effective exponent are
in good agreement with the experimental data at low $x$ values. We
have studied the effects of adding the nonlinear GLR-MQ
corrections to the NNLO linear behavior of the gluon distribution
function and reduced cross section. These results are  close to
the data for low-$Q^{2}$ values as we have discussed the meaning
of these findings from the points of view of modified nonlinear
and higher twist corrections added to the structure function.\\

\subsection{Acknowledgments}
The authors appreciate the Research Council of Razi University for
official support of this work. G.R.Boroun thanks the Department of
Physics of the CERN-TH for their warm hospitality. Also authors
would like to thank H.Khanpour for discussions which completed
this study.\\


\subsection{Appendix A}

The gluon distribution is derived from the leading order DGLAP
evolution equation for $F_{2}(x,Q^{2})$ in a domain $x{\leq}0.09$
and $0.11~\mathrm{GeV^{2}}{\leq}Q^{2}{\leq}1200~\mathrm{GeV^{2}}$.
Solution for $G(x,Q^{2})$ using a Froissart bounded structure
function $F_{2}(x,Q^{2})$ is obtained in terms of a quadratic
polynomial in ${\ln}(1/x)$ with quadratic polynomial coefficients
in ${\ln}(Q^{2})$ as [12]
\begin{eqnarray}
G(x,Q^{2})=-0.459-0.143{\ln}(Q^{2})-0.0155{\ln}^{2}(Q^{2})\nonumber\\
+[0.231+0.00971{\ln}(Q^{2})-0.0147{\ln}^{2}(Q^{2})]{\ln}(1/x)\nonumber\\
+[0.0836+0.06328{\ln}(Q^{2})+0.0112{\ln}^{2}(Q^{2})]{\ln}^{2}(1/x).
\end{eqnarray}
The above method generalized to the case of massive quarks
directly from $F_{2}(x,Q^{2})$ by Laplace transforms to the DGLAP
equation [13]. The gluon distribution function for $n_{f}=4$ at
$Q^{2}=M^{2}_{b}$ with $M_{b}=4.5~\mathrm{GeV}$ is obtained with
an expression quadratic in both ${\ln}(1/x)$ and ${\ln}(Q^{2})$ by
the following form
\begin{eqnarray}
G_{4}(x,Q^{2})=\frac{3}{5}\mathcal{H}_{3}(x,Q^{2})=\mathcal{H}_{4}(x,Q^{2}),
\end{eqnarray}
where
\begin{eqnarray}
\mathcal{H}_{3}(x,Q^{2})=-2.94-0.359{\ln}(Q^{2})-0.101{\ln}^{2}(Q^{2})\nonumber\\
+[0.594-0.0792{\ln}(Q^{2})-0.000578{\ln}^{2}(Q^{2})]{\ln}(1/x)\nonumber\\
+[0.168+0.138{\ln}(Q^{2})+0.0169{\ln}^{2}(Q^{2})]{\ln}^{2}(1/x).
\end{eqnarray}


\newpage{
\section{References}

1. M. Klein, Ann. Phys.{\bf528}, 138 (2016).\\
2. R.A.Khalek et al., arXiv:1906.10127 [2019].\\
3. A. Abada et al., [FCC Collaborations], Eur.Phys.J.C{\bf 79}, 474(2019).\\
4. A.Donnachie and P.V.Landshoff, Phys.Lett.B {\bf470}, 273(1999).\\
5. M.Froissart, Phys.Rev.{\bf123}, 1053 (1961).\\
6. Yu.L.Dokshitzer, Sov.Phys.JETP {\textbf{46}}, 641(1977);
G.Altarelli and G.Parisi, Nucl.Phys.B \textbf{126}, 298(1977);
V.N.Gribov and L.N.Lipatov, Sov.J.Nucl.Phys. \textbf{15},
438(1972).\\
7. H.Khanpour et al., Phys.Rev.C{\bf95}, 035201(2017); H.Khanpour, Phys.Rev. D{\bf99}, 054007(2019).\\
8. E. L. Berger, M. M. Block and C. I. Tan, Phys. Rev. Lett. {\bf
98}, 242001 (2007); M. M. Block, E. L. Berger and C. I. Tan, Phys.
Rev. Lett.{\bf 97}, 252003 (2006).\\
9. M. M. Block, L. Durand, P. Ha and D. W. McKay, Phys. Rev.{\bf D
84},
094010 (2011); Phys. Rev.{\bf D 88}, no. 1, 014006 (2013).\\
10. M. M. Block, L. Durand and P. Ha, Phys. Rev.{\bf D 89}, no. 9,
094027 (2014).\\
11. F. D. Aaron et al. [H1 and ZEUS Collaborations], JHEP{\bf
1001}, 109 (2010).\\
12.  M. M. Block, L. Durand and D. W. McKay, Phys. Rev.{\bf D 77},
094003 (2008); Phys. Rev.{\bf D 79}, 014031 (2009).\\
13. M.M.Block, Eur.Phys.J.{\bf C65}, 1(2010); M. M. Block and  L.
Durand, arXiv:0902.0372[hep-ph](2009).\\
14. K.Prytz, Phys.Lett.{\bf B311},286(1993); A.V.Kotikov and
G.Parente, Phys.Lett.{\bf B379}, 195(1996); G.R.Boroun and
B.Rezaei, Chin.Phys.Lett.{\bf 23}, 324(2006).\\
15. G.R.Boroun and B.Rezaei, Eur.Phys.J.C{\bf73},
2412(2013); S.Zarrin and G.R.Boroun, Nucl.Phys.{\bf B922}, 126 (2017);
M.Lalung et al., Int.J.Theor.Phys.{\bf 56}, 3625(2017); M.Devee, arXiv:1808.00899(2018).\\
16. A.V.Kotikov, Phys.Atom.Nucl,{\bf 80}, 572 (2017).\\
17. N.Yu.Chernikova and A.V.Kotikov, JETP Lett.{\bf 105}, 223(2017).\\
18. A.Donnachie and P.V.Landshoff, Phys.Lett.B {\bf550}, 160(2002 ).\\
19. W.L. van Neerven, A.Vogt, Phys.Lett.B \textbf{490}, 111(2000).\\
20. A.Vogt, S.Moch, J.A.M.Vermaseren, Nucl.Phys.B \textbf{691}, 129(2004).\\
21. B.G. Shaikhatdenov, A.V. Kotikov, V.G. Krivokhizhin, G.
Parente, Phys. Rev. D {\bf81}, 034008(2010).\\
22. K Golec-Biernat and A.M.Stasto, Phys.Rev.D {\bf80},
014006(2009); R.D.Ball et al., Eur.Phys.J. C{\bf76}, 383(2016).\\
23. B.Rezaei and G.R.Boroun, Eur.Phys.J.A{\bf55}, 66(2019); G.R.Boroun, Eur.Phys.J.A{\bf50}, 69(2014).\\
24. L.A.Anchordoqui et al., arXiv:1902.10134[hep-ph].\\
25. G.Altarelli and G.Martinelli, Phys.Lett.B\textbf{76}, 89(1978).\\
26. S.Moch, J.A.M.Vermaseren and A.Vogt, Phys.Lett.{\bf B606},
123(2005).\\
27. A. H. Mueller and J. Qiu, Nucl. Phys. B\textbf{268}(1986)427;
L. V. Gribov, E. M. Levin and M. G. Ryskin, Phys.
Rep.\textbf{100},
 (1983)1.\\
28. G. R. Boroun and B. Rezaei, Chin. Phys. Lett.{\bf32}, (2015)
no.11, 111101; B. Rezaei and G. R. Boroun, Phys. Lett. B{\bf692}
(2010) 247;
 G. R. Boroun, Eur. Phys. J. A{\bf43} (2010) 335.\\
29. N. N. Nikolaev and W. Sch$\ddot{a}$fer, Phys. Rev. D{\bf74}(2006)014023.\\
30. K. J. Eskola et al., Nucl. Phys. B{\bf660}(2003)211.\\
31. R. Fiore, P. V. Sasorov and V. R. Zoller, JETP Letters
{\bf96}(2013)687; R. Fiore, N. N. Nikolaev and V. R. Zoller,
JETP Letters {\bf99}(2014)363.\\
32. A.M.Cooper-Sarkar, arXiv:1605.08577v1 [hep-ph] 27 May 2016; I.Abt et.al., arXiv:1604.02299v2 [hep-ph] 11 Oct 2016.\\
33. F.D. Aaron et al. [H1 Collaboration], Eur.Phys.J. C{\bf63}, 625(2009).\\
34. K.Prytz, Phys.Lett.B{\bf311}, 286(1993); Phys.Lett.B{\bf332},
393(1994).\\
35. M.B.Gay Ducati and P.B.Goncalves, Phys.Lett.B{\bf390},
401(1997).\\
36. G.R.Boroun, Journal of Experimental and Theoretical Physics,
{\bf111}, 566(2010); N.N.Nikolaev and B.G.Zakharov, Phys.Lett. B{\bf332}, 184(1994).\\
37.  N.N.Nikolaev and B.G.Zakharov, Phys.Lett.B332, 184(1994);
N.N.Nikolaev and V.R.Zoller, Phys.Atom.Nucl.73, 672 (2010).\\
38. P.Jimenez-Delgado and E.Reya, Phys.Rev.{\bf D79}, 074023(2009).\\
39. C.Adloff et al., [H1 Collaboration], Eur.Phys.J.C.{\bf 21},
33(2001).\\
40. F.D. Aaron et al., [H1 Collaboration], Phys.Lett.B.{\bf 665},
139(2008).\\
 }

\begin{figure}
\includegraphics[width=0.55\textwidth]{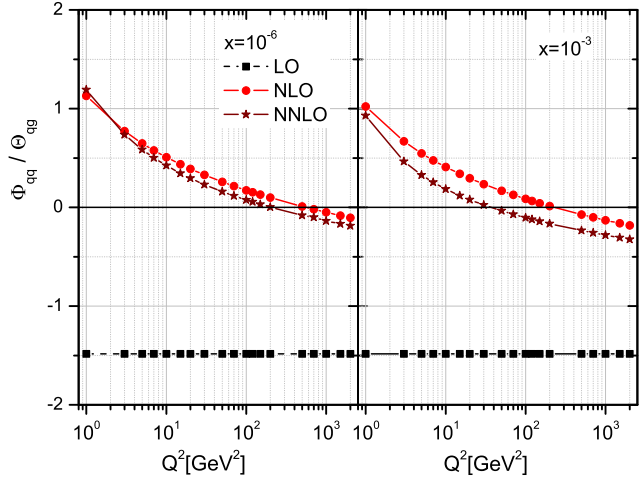}
\caption{Plot of the ratio $\frac{\Phi_{qq}}{\Theta_{qg}}$ as a
function of $Q^{2}$ for two $x$ ($1\times 10^{-6}$ and $1\times
10^{-3}$).}\label{Fig1}
\end{figure}
\begin{figure}
\includegraphics[width=0.55\textwidth]{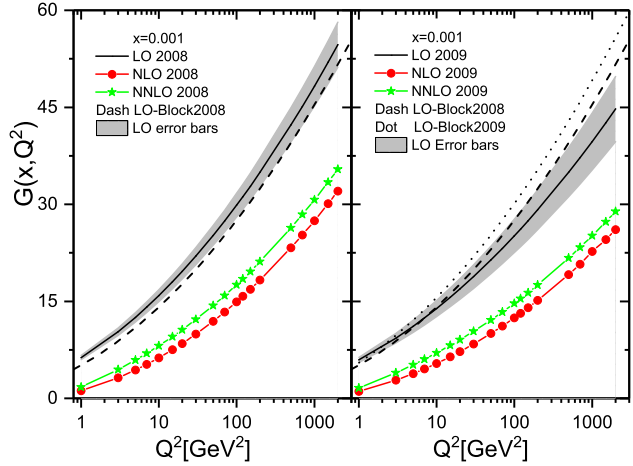}
\caption{The leading-order up to the next-to next-to leading order
gluon distributions in comparison with the leading-order results
of Block model [12-13]. The error bars are due to the $F_{2}$
parameterization at LO approximation.}\label{Fig2}
\end{figure}
\begin{figure}
\includegraphics[width=0.55\textwidth]{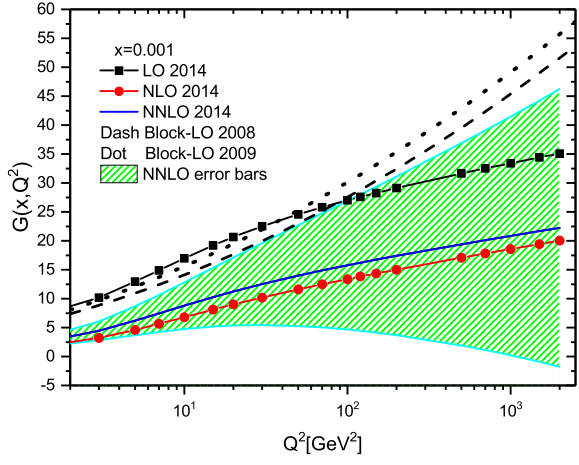}
\caption{The leading-order up to the next-to next-to leading order
gluon distributions from the new $F_{2}$ parameterization [10]
compared with the leading-order results of Block model [12-13].
The error bars are due to the $F_{2}$ parameterization at NNLO
approximation. }\label{Fig3}
\end{figure}
\begin{figure}
\includegraphics[width=1\textwidth]{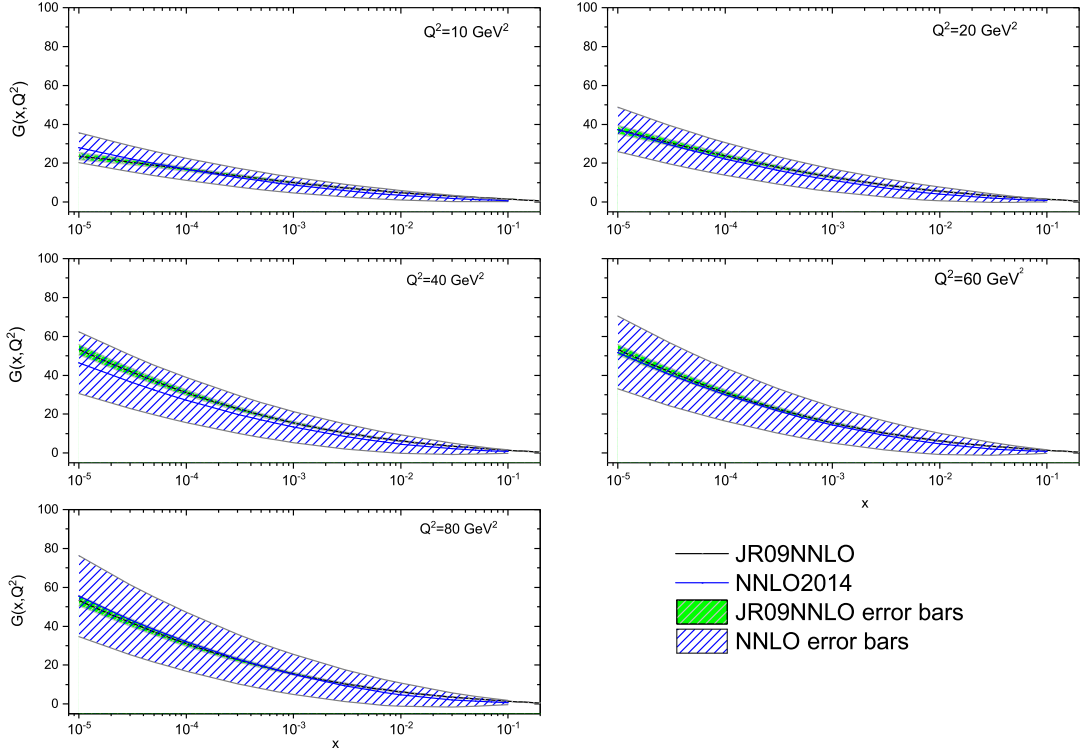}
\caption{The NNLO gluon distributions from the new $F_{2}$
parameterization [10] and its comparison with the results from
 JR09NNLO global QCD analysis [38] at $Q^{2}=10, 20, 40, 60 $ and
$80~\mathrm{GeV}^{2}$. The error bars are due to the $F_{2}$
parameterization at NNLO approximation and JR09 NNLO
analysis.}\label{Fig4}
\end{figure}
\begin{figure}
\includegraphics[width=0.55\textwidth]{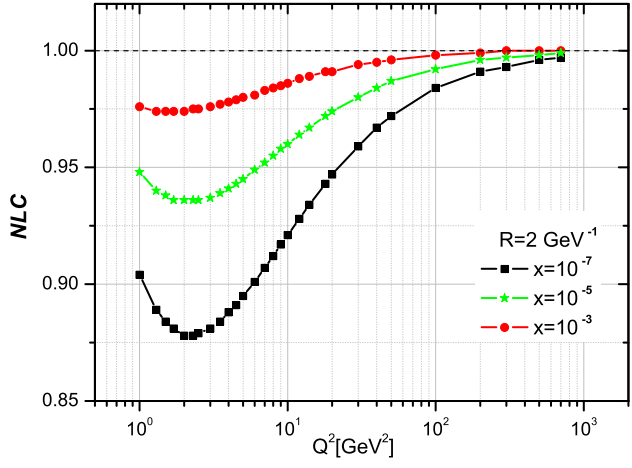}
\caption{Nonlinear correction (NLC) effects for
$R=2~\mathrm{GeV}^{-1}$ in a wide range of $Q^{2}$ at low and very
low $x$ values.}\label{Fig4}
\end{figure}
\begin{figure}
\includegraphics[width=0.55\textwidth]{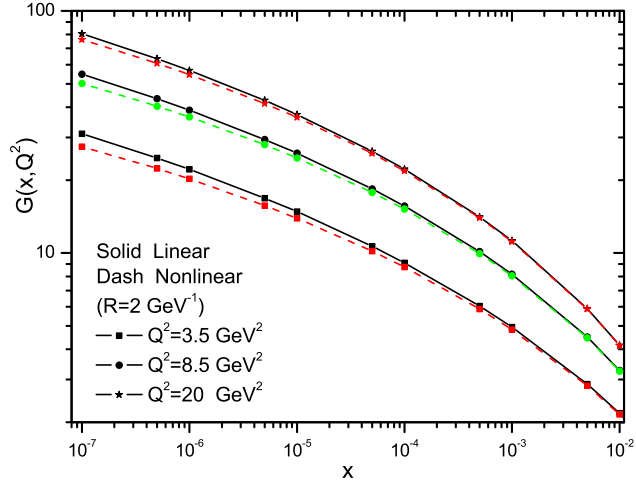}
\caption{x-evolution of linear and nonlinear gluon distribution
function at NNLO approximation for three fixed
$Q^{2}$.}\label{Fig4}
\end{figure}
\begin{figure}
\includegraphics[width=0.6\textwidth]{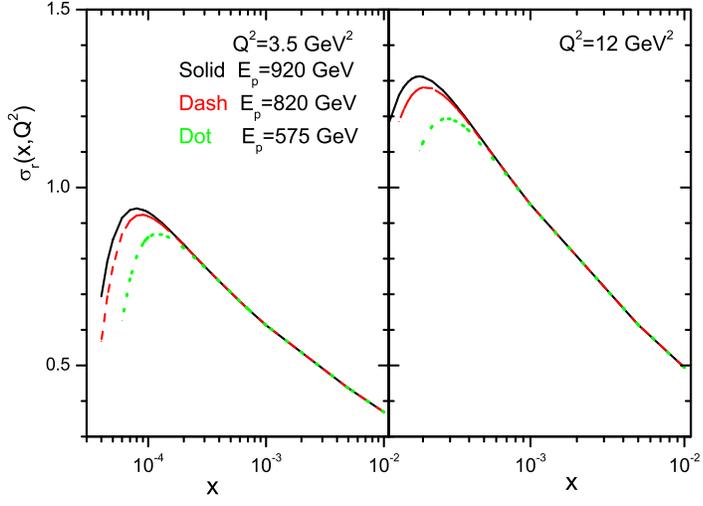}
\caption{The NNLO predictions for $\sigma_{r}(x,Q^{2})$ but for
different proton beam energies $E_{p}$ at $Q^{2}=3.5$ and
$12~\mathrm{GeV}^{2}$. Notice that the curves terminate when
$y=1$.}\label{Fig7}
\end{figure}
\begin{figure}
\includegraphics[width=1\textwidth]{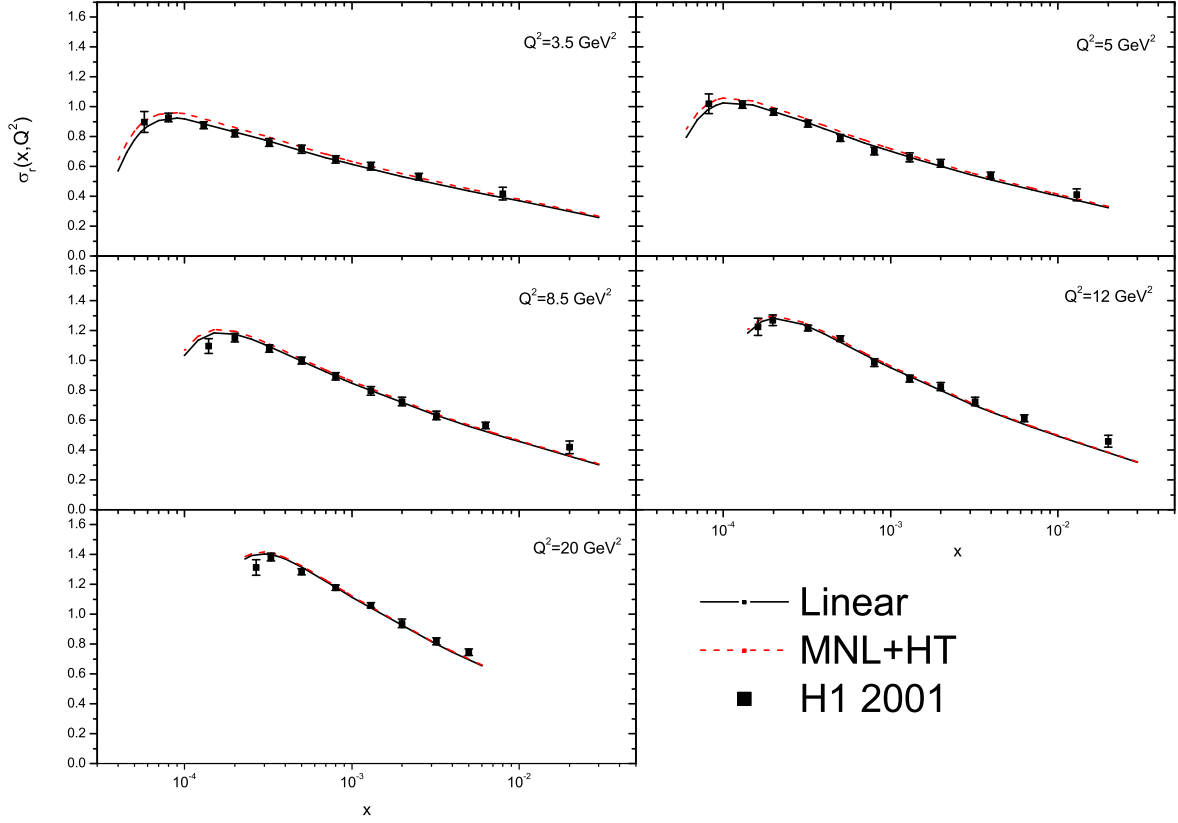}
\caption{The NNLO predictions for the reduced DIS cross section.
The modified nonlinear (MNL) and higher twist (HT) corrections
represent our results at NNLO. The H1 data for some representative
fixed values of $Q^{2}$ are taken from [39].}\label{Fig8}
\end{figure}
\begin{figure}
\includegraphics[width=0.6\textwidth]{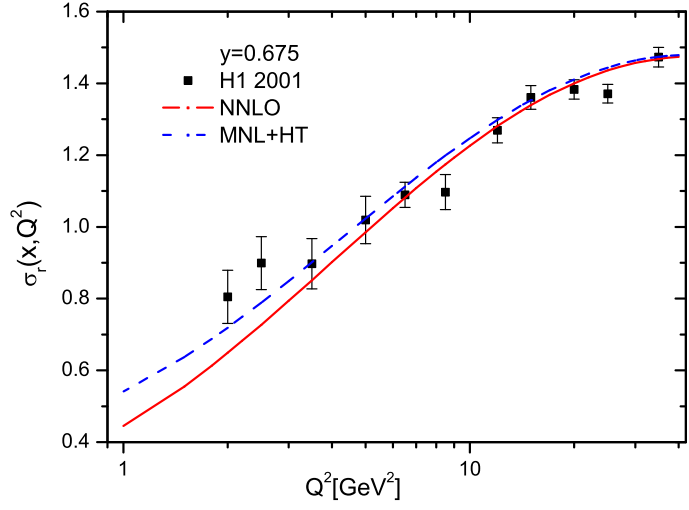}
\caption{The NNLO predictions for the reduced DIS cross section at
high inelasticity available. The H1 data for some representative
fixed values of $Q^{2}$ are taken from [39] as accompanied with
total errors. Modified nonlinear and higher twist corrections
compared with linear results in a wide range of $Q^{2}$.
}\label{Fig9}
\end{figure}

\end{document}